\documentclass[preprint,aps,nofootinbib,tightenlines,showpacs]{revtex4}
\usepackage{amsmath}
\usepackage{amsfonts}
\usepackage{amssymb}
\usepackage{latexsym}
\usepackage{graphicx}
\usepackage{bm}

\def\nn{\nonumber}
\def\dx{dx}
\def\dt{dt}

\def\dps{\displaystyle}
\def\hox{\left(\frac{x}{2}\right)}

\def\ra{\rightarrow}
\def\lra{\longrightarrow}
\def\re{\mathrm{Re}}

\def\half{\frac{1}{2}}
\def\bx{\mathbf{x}}\def\bk{\mathbf{k}}\def\bp{\mathbf{p}}
\def\CH{\mathcal{H}}

\def\CC{\mathcal{C}}

\def\M{M_\mathrm{eff}}
\def\ep{\epsilon}
\def\sk{\sum^\infty_{k=0}\frac{(-)^k}{k!}}
\usepackage{color}

\begin{document}
\newcommand{\be}{\begin{equation}}
\newcommand{\ee}{\end{equation}}
\newcommand{\bea}{\begin{eqnarray}}
\newcommand{\eea}{\end{eqnarray}}
\newcommand{\barr}{\begin{array}}
\newcommand{\earr}{\end{array}}


\title{Curvature Perturbation Spectrum in Two-field Inflation with
 a Turning Trajectory}
\author{Shi Pi}\email{spi@pku.edu.cn}\affiliation{Department of Physics and
\\ State Key Laboratory of Nuclear Physics and Technology,
Peking University, Beijing 100871,
China}
\author{Misao Sasaki}\email{misao@yukawa.kyoto-u.ac.jp}
\affiliation{Yukawa Institute for Theoretical Physics, Kyoto University,
Kyoto 606-8502, Japan}
\begin{abstract}
We revisit a two-component inflaton model with a turning trajectory in the
field space, where the field slowly rolls down along the trajectory.
We consider the case
when the effective mass in the direction perpendicular to the trajectory,
namely the isocurvature direction, is either of the same order as or much
larger than the Hubble parameter. Assuming that the turning angular
velocity is small, we compute analytically the
corrections to the power spectrum of
curvature perturbation caused by the mediation of the heavy isocurvature
perturbation, and compare our analytic results with the numerical ones.
Especially, when $M_{\mathrm{eff}}^2\gg H^2$, we
find that it is proportional to $M_\mathrm{eff}^{-2}$.
This result is consistent with the one obtained previously by an
effective field theory approach.
\end{abstract}

\preprint{YITP-12-36}

\maketitle



\section{Introduction}\label{Sec:intro}
Inflation~\cite{inflation_bible,newinflation} has been recognized as
the most competitive
 model for the early universe, which can not only solve basic problems
 of big-bang cosmology, say, the horizon and flatness problems
and overproduction of topological defects, but also give a natural
explanation for the primordial fluctuations that account for
the observed CMB anisotropy and large scale structure of the universe.

The simplest model of inflation is driven by a single scalar field with
a canonical kinetic term and sufficiently flat potential, which generates
an almost scale-invariant, highly Gaussian curvature
perturbation~\cite{Mukhanov:1982nu}.
But one can still consider a single-field model with complex kinetic term like
k-inflation~\cite{ArmendarizPicon:1999rj}
or DBI inflation~\cite{Silverstein:2003hf}.
The WMAP has accurately
confirmed the existence of primordial curvature perturbations
with a nearly scale-invariant spectrum of order $10^{-9}$, but
at the same time indicated the possible existence of non-Gaussianity that
cannot be explained by a single-field slow-roll inflation with canonical
kinetic term~\cite{Komatsu:2010fb}.

Multi-field inflation is another natural generalization of
the simplest single-field inflation.
It is motivated partly because of some
theoretical considerations, like string landscape~\cite{Susskind:2003kw},
 and partly because it can lead to more abundant phenomena.
For example, N-flation or assisted inflation~\cite{Liddle:1998jc}
shows that $N$ scalar fields move collectively
to sustain inflation to an e-folding number required, even if each
individual field is unable to drive inflation.
Also one can consider multi-field k-inflation or DBI
inflation~\cite{Huang:2007hh,Ward:2007gs,Pi:2011tv}.

As a case study, two-field inflation can bring us fruitful new properties while
maintaining geometrical intuitions in the field space.
In slow-roll paradigm, we can classify two-field models into
different categories by their masses, velocity in field
space, etc.. The simplest case is when both fields are almost massless.
In this case the effect of the curvature of the trajectory in field space
can be neglected, implying that the equations of motion can be
effectively decomposed into the one along the trajectory and the other
orthogonal to the trajectory.
One then call the field along the trajectory the ``adiabatic'' component
and the one orthogonal to it the ``entropy'' or ``isocurvature'' component.
The curvature perturbation (literally speaking the curvature perturbation on
comoving or uniform density slices) is directly related to
the adiabatic component of the field perturbation (on flat slices),
and is affected by the entropy perturbation via ``transfer function'',
which can be calculated in super-horizon
era~\cite{Sasaki:1998ug,Gordon:2000hv,Amendola:2001ni,Peterson:2010np}.
Or, one can calculate directly the power spectrum by using the $\delta N$
formalism~\cite{Sasaki:1995aw,Lyth:2004gb,Starobinsky:1986fxa,Starobinsky:1982ee}.

Another important case is when the trajectory is along a deep
valley, that is, the effective mass along the trajectory is sufficiently
small so that the field is in slow-roll motion, while the effective
isocurvature mass perpendicular to the trajectory is heavy, $M_\mathrm{eff}\gg H$.
In such a model, the behavior is more or less like that in a single-field
theory since the isocurvature perturbation quickly decays out.
Namely, integrating out the heavy field, the system can be described
by an effective single field
theory.
In this effective field approach, it has been recently
claimed that the effect of the heavy field can be
absorbed into a correction in the sound speed which is
inversely proportional to the squared effective mass of
the isocurvature component~\cite{Tolley:2009fg,Cremonini:2010ua,Achucarro:2010da}.
However, because of some non-trivial assumptions in this approach
it is desirable to compute the correction
without resorting to the effective field approach, and justify
or falsify it.

The case that adjoins these models is called ``quasi-single field''
inflation by~\cite{Chen:2009zp}, in which the effective mass of
isocurvature mode $M_\mathrm{eff}$ is of the same order as the Hubble
 parameter $H$.
Assuming for simplicity that the field is approximately in a circular
motion with a small and constant angular velocity, one can use the
 in-in formulism~\cite{inin,Weinberg:2005vy,Maldacena:2002vr} to calculate
 the 2-point function of the curvature perturbation by quantizing the
 curvature and isocurvature perturbations as free fields, with the
coupling terms as interaction vertices~\cite{Chen:2009we,Gao:2009qy}.
They calculated the correction to power spectrum in
$M_\mathrm{eff}\sim H$ case, and found it proportional to the angular
 velocity squared with a $M_\mathrm{eff}$-dependent coefficient of
order 1, which was estimated by numerical methods~\cite{Chen:2009we}.

The aim of the current paper is to apply the method developed
in the ``constant turn'' quasi-single field inflation in~\cite{Chen:2009zp}
to calculate analytically the coefficient of the corrections caused by the
mediation of isocurvaton to the power spectrum. Especially for the case when
the effective mass of the isocurvature
mode is large, $M_\mathrm{eff}^2\gg H^2$, we examine the result
obtained by the effective field theory
approach~~\cite{Tolley:2009fg,Achucarro:2010da},
which is found to be
consistent with our analytic result.

This paper is organized as follows. In Section~\ref{Sec:QSFI} we briefly
 review the quasi-single field inflation and the in-in formulism used to
 calculate the power spectrum of curvature perturbation,
 and calculate its coefficient analytically .
In Section~\ref{Sec:EFTI} we extend our calculation to the case of a
heavy isocurvature mode, and compare our result with the one obtained
by a totally different method, namely, the effective field approach
in which the heavy field is integrated out from the beginning.
We conclude our paper in Section~\ref{Sec:conclude}.
Some detailed calculations are spelled out in Appendix.

\section{Light Isocurvature Modes: Quasi-Single Field Inflation}\label{Sec:QSFI}
In this section we give a brief review of a simple two-field inflation
model: the quasi-single field inflation. Using the in-in formalism,
we derive corrections to the power spectrum of the curvature perturbation
due to interactions with the isocurvature mode.

Quasi-single field inflation describes a segment of time
during generic inflation when the curvature field is massless and undergoes
 a slow-roll trajectory, while the isocurvature mode has mass of order $H$.
 To be specific, we consider a motion along an arc with radius $R$.
Then we can naturally decompose the field into the curvature mode
along the tangent of the circle, $R\theta$, and the isocurvature mode
given by the radial field $\sigma$. We call $\sigma$ the isocurvaton.
The Lagrangian of such a system is
\bea
S_m = \int d^4x \sqrt{-g} \left[
-\half(\tilde R+\sigma)^2 g^{\mu\nu} \partial_\mu \theta \partial_\nu \theta
- \half g^{\mu \nu} \partial_\mu \sigma \partial_\nu \sigma
- V_{\rm sr}(\theta) - V(\sigma) \right],
\label{ModelAction}
\eea
where  $V_{\rm sr}(\theta)$ is a usual slow-roll potential,
$V(\sigma)$ is a potential that forms a circular valley and
traps the isocurvaton at $\sigma=\sigma_0$.

The equations of motion for classical trajectory is then
\begin{eqnarray}
     3M_p^2 H^2&=&\frac{1}{2}R^2\dot\theta_0^2+V+V_{\rm sr},\\
    -2M_p^2 \dot H&=& R^2\dot\theta_0^2 ,\\
    0&=&R^2\ddot\theta_0+3R^2H\dot\theta_0+V_{\rm sr}',\\
    0&=&\ddot\sigma_0+3H\dot\sigma_0+V'-R^2\dot\theta_0^2\,,
\end{eqnarray}
where $R\equiv \tilde{R}+\sigma_0$. Now $\dot\sigma_0=0$ gives us
 $V'(\sigma_0)=R\dot\theta_0^2$.
This relation shows that the field is not exactly at the
bottom of the valley of the potential $V$, but is slightly shifted
from it to provide the centripetal force of circular motion.
 In what follows we omit the argument $\sigma_0$ of $V'$, $V''$, $V'''$, etc.,
unless there is a chance of confusion. We denote the
``rotation speed'' by $\dot\theta$ as in~\cite{Gordon:2000hv}
with a minus sign because of the different orientation.
In the effective-field-theory
 papers~\cite{Achucarro:2010da,Cespedes:2012hu,Achucarro:2012sm},
they used $\eta_\perp=\dot\theta/H$ instead.
We define the slow-roll parameters along the trajectory as
\begin{align}
\epsilon &\equiv - \frac{\dot H}{H^2}
= \frac{R^2\dot\theta_0^2}{2H^2M_\mathrm{Pl}^2}
\approx \frac{M_\mathrm{Pl}^2}{2}
\left( \frac{V_\mathrm{sr}'}{R V_\mathrm{sr}}\right)^2,
\cr
\eta &\equiv \frac{\dot\epsilon}{H\epsilon}
\approx -2 M_\mathrm{Pl}^2 \frac{V_\mathrm{sr}''}{R^2 V_\mathrm{sr}}
+ 2M_\mathrm{Pl}^2\left( \frac{V_\mathrm{sr}'}{R V_\mathrm{sr}}\right)^2.
\end{align}
We will not define the ``slow-roll parameters'' perpendicular to the
trajectory since the field does not slowly roll in this direction,
and since they can be represent by $\dot\theta_0$ and $V''$.

\begin{figure}
  \begin{minipage}[t]{0.4\linewidth}
    \centering
    \includegraphics[width=0.8\textwidth]{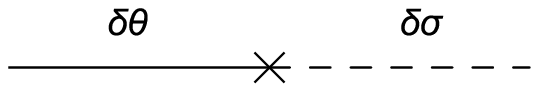}
    \caption{The second order interacting vertex $\CH_2$
 which is proportional to $\dot\theta/H$.}
    \label{Fig:vertex}
  \end{minipage}%
  ~~~~~~
  \begin{minipage}[t]{0.4\linewidth}
    \centering
    \includegraphics[width=0.8\textwidth]{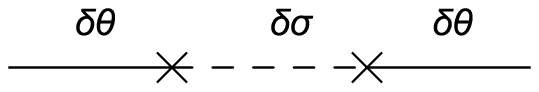}
    \caption{The leading order to the 2-point function of curvature
perturbation, mediated by a isocurvature field.}
    \label{Fig:2p}
  \end{minipage}
\end{figure}

Now let us consider the perturbation to $\theta$ and $\sigma$
in the spatially flat gauge where
\bea
h_{ij}= a^2(t) \delta_{ij},
\label{gauge1metric}
\eea
and
\bea
\theta(t,\bx) = \theta_0(t) + \delta\theta(t,\bx) , ~~~~~
\sigma(t,\bx) = \sigma_0 + \delta\sigma(t,\bx).
\label{gauge1fields}
\eea
In this gauge in the leading order in the slow-roll approximation,
the gravitational effect can be neglected, and we obtain
the Hamiltonian density,
\bea \label{H0}
\CH_0 &=& a^3 \left[ \half R^2 \dot {\delta\theta_I}^2 +
  \frac{R^2}{2a^2}
  (\partial_i \delta\theta)^2
+ \half \dot{\delta\sigma}^2 + \frac{1}{2a^2} (\partial_i
\delta\sigma)^2 + \half \M^2 \delta \sigma^2
\right],\\\label{CH2}
\CH^I_2 &=& -c_2 a^3 \delta\sigma \dot{\delta\theta},
\qquad c_2 = 2 R \dot\theta,
\\\label{CH3}
\CH^I_3 &=&  -a^3R\delta\sigma\dot{\delta\theta}^2
-a^3\dot\theta\dot{\delta\theta}\delta\sigma^2
+aR\delta\sigma\left(\partial_i\delta\theta\right)^2
+\frac{a^3}{6}V'''\delta\sigma^3,\\
\M^2&=& V'' + 3\dot \theta^2,
\eea
where and below we omit the subscript $0$ from the background
quantities. It is known that the conserved curvature perturbation on
comoving slices ${\cal R}_c$ is given in terms of the field fluctuation
in the flat slices along the trajectory $\delta\theta$
as~\cite{Sasaki:1986hm,Mukhanov:1988jd}
\begin{eqnarray}
{\cal R}_c=-\frac{H\delta\theta}{\dot\theta}\,.
\label{Rcform}
\end{eqnarray}
For the ``constant turn'' case, $c_2$ and $\M^2$ are both constants.
$\M$ is the effective mass of the isocurvature field. We split the
Hamiltonian into the free part $\CH_0$ and interacting parts
$\CH_2$, $\CH_3$. Next we will treat $\CH_2$ as interacting vertex
to the free Hamiltonian $\CH_0$ to calculate the power spectrum,
 while this vertex is proportional to $\dot\theta_0/H$ and depicted
in Fig.~\ref{Fig:vertex}. $\CH_3$ is only useful when turning to
investigate the bispectrum which is not investigated in the present paper.
This method is valid only when the interacting Hamiltonian is small, i.e.,
\begin{equation}
  \left(\frac{\dot\theta}{H}\right)^2 \ll 1\,,
  \quad \frac{|V'''|}{H} \ll 1\,.
\end{equation}
Then the perturbation theory can be used, and there is no constraint
on the value of $\M\sim V''$: It maybe small, of order $H$, or large.
 The terminology ``quasi-single field'' denotes the case
$\M\sim H$ which we review in this section.
The case when $\M\gg H$ will be studied in Section~\ref{Sec:EFTI}
 by the same approach.

In the interaction picture, we quantize the Fourier components $\delta
\theta_\bk^I$ and $\delta\sigma_\bk^I$ of the free fields
$\delta\theta^I$ and $\delta\sigma^I$,
\bea
\delta\theta_\bk^I &=& u_\bk a_\bk + u_{-\bk}^* a_{-\bk}^\dagger,
\\
\delta\sigma_\bk^I &=& v_\bk b_\bk + v_{-\bk}^* b_{-\bk}^\dagger,
\eea
where $a_\bk$ and $b_\bk$ are the annihilation operators
of $\delta\sigma$ and $\delta\theta$, respectively, and each one
satisfies the canonical commutation relation,
\bea
[a_\bk,a_{-\bk'}^\dagger] = (2\pi)^3 \delta^3 (\bk+\bk'),
\quad
[b_\bk,b_{-\bk'}^\dagger] = (2\pi)^3 \delta^3 (\bk+\bk'),
\eea
with all the other commutators equal to zero.
The mode functions, $u_\bk$ and $v_\bk$, satisfy the linear equations of
motion followed from the Hamiltonian $\CH_0$,
\bea
u_\bk'' - \frac{2}{\tau} u_\bk' + k^2 u_\bk &=&0,
\label{modefun_u} \\
v_\bk'' - \frac{2}{\tau} v_\bk' + k^2 v_\bk +
\frac{\M^2}{H^2 \tau^2} v_\bk &=& 0,
\label{modefun_v}
\eea
where $\tau$ is the conformal time defined by
$\mathrm{d}\tau=\mathrm{d}t/a(t)$, the prime
denotes the derivative with respect to $\tau$, and
we have approximated the background by a de Sitter spacetime.

The solutions to (\ref{modefun_u}) and (\ref{modefun_v}) are
given by linear combinations of the Hankel functions of first and second
kind. Requiring that the solutions approach those in the Minkowski
positive frequency functions, or those for
the Bunch-Davies vacuum~\cite{Bunch:1978yq}, in the limit $k\gg aH$,
\bea
R u_\bk\,,\quad
v_\bk \to i \frac{H}{\sqrt{2k}} \tau e^{-i k\tau},
\eea
we obtain
\bea
u_\bk = \frac{H}{R\sqrt{2k^3}} ( 1+i k\tau)e^{-i k\tau},
\label{mode_u}
\eea
and
\bea
v_\bk = -i e^{i(\nu+\half)\frac{\pi}{2}} \frac{\sqrt{\pi}}{2} H
  (-\tau)^{3/2} H^{(1)}_\nu (-k\tau),
\quad {\rm for}~~ \M^2/H^2 \le 9/4,
\label{mode_v1}
\eea
where $\nu = \sqrt{9/4-\M^2/H^2}$, or
\bea
v_\bk = -i e^{ -\frac{\pi}{2}\mu + i\frac{\pi}{4}}
\frac{\sqrt{\pi}}{2} H
  (-\tau)^{3/2} H^{(1)}_{i\mu} (-k\tau),
\quad {\rm for}~~ \M^2/H^2 > 9/4,
\label{mode_v2}
\eea
where $\mu = \sqrt{\M^2/H^2-9/4}$.

Next we consider the interaction terms as perturbations to
 $\CH_0$. To calculate the power spectrum up to tree-level diagrams,
we just need to consider the $\CH_2$ term which describes the correlation
between the curvature and isocurvature modes,
\bea
H^I_2 = \int \mathrm{d}^3 \bx ~\CH^I_2
= -c_2 a^3 \int \frac{\mathrm{d}^3\bk}{(2\pi)^3}
\delta\sigma^I_\bk \dot{\delta\theta}^I_{-\bk} ~.
\eea
The two-point function of $\delta\theta^2$ can be calculated
by the in-in formulism as
\bea
\langle \delta\theta^2 \rangle &\equiv&
\langle 0| \left[ \bar T \exp\left( i\int_{t_0}^t \dt' H_I(t')\right)
  \right] \delta\theta_I^2(t)
  \left[T \exp\left( -i\int_{t_0}^t \dt' H_I(t')\right)
  \right] |0\rangle,
\\\nn
&=&\langle0|\bar T\left(1+i\int_{t_0}^t \dt'H_I(t')
-\int_{t_0}^t\dt_1\int_{t_0}^t\dt_2H_I(t_1)H_I(t_2)+...\right)
  \delta\theta_I^2(t)\\
&&  \cdot~T\left(1+i\int_{t_0}^t \dt'H_I(t')
-\int_{t_0}^t\dt_1\int_{t_0}^t\dt_2H_I(t_1)H_I(t_2)+...\right)|0\rangle,\\
&=&\langle 0| \delta \theta_I^2 |0\rangle
\cr
&+& \int_{t_0}^t \dt_1 \int_{t_0}^t \dt_2
\langle 0| H_I(t_1) ~\delta\theta_I^2~ H_I(t_2) |0\rangle
\label{Original A}\\
&-& 2 ~{\rm Re} \left[
\int_{t_0}^t dt_1 \int_{t_0}^{t_1} dt_2
\langle 0| \delta\theta_I^2~ H_I(t_1) H_I(t_2) |0\rangle \right]
\label{Original B} \\
&+& \cdots ~.
\nonumber
\eea

To calculate the terms sandwiched with the free field vacuum, we use
the normal ordering. After that only fully contracted terms survive.
The correction from a mediating isocurvature perturbation is
depicted in Fig.~\ref{Fig:2p}.
This gives the correction to the power spectrum of $\delta\theta$ as
\begin{align}
 (2\pi)^3 \delta^3 (\bp_1+\bp_2) \frac{c_2^2}{R^4}
 \frac{\cal C(\nu)}{p_1^3}~,
\end{align}
with a factor $\mathcal{C}(\nu)$ which only depends on the effective
mass $M_\mathrm{eff}$ and rotation speed $\dot\theta^2$ via $\nu$,
\begin{eqnarray}\label{nu and m}
\nu&=&\sqrt{\frac{9}{4}-\frac{M_\mathrm{eff}^2}{H^2}},\;\;\;\;M_\mathrm{eff}^2
=V_{\sigma\sigma}+3\dot\theta^2,\\\nn
\mathcal{C}(\nu)&=&
\frac{\pi}{8}\left\{\left|\int_0^\infty \dx x^{-1/2}H_\nu^{(1)}(x)e^{ix}\right|^2\right.
\\
&-&\left.2\mathrm{Re}\int_0^\infty \dx_1x_1^{-1/2}H_\nu^{(1)}(x_1)e^{-ix_1}
\int_{x_1}^\infty \dx_2x_2^{-1/2}\left(H_\nu^{(1)}(x_2)\right)^\ast e^{-ix_2}\right\}\,.
\label{C(nu)}
\end{eqnarray}
Applying the above to Eq.~(\ref{Rcform}), the power spectrum of
the curvature perturbation on comoving slices is given by
\begin{equation}\label{P-C}
{\cal P}_{\cal R}=\frac{H^4}{4\pi^2R^2\dot\theta^2}
\left[1+8\mathcal{C}(\nu)\left(\frac{\dot\theta}{H}\right)^2\right].
\end{equation}
The calculation of $\mathcal C(\nu)$ for real $\nu$ has been done
in~\cite{Chen:2009zp} by numerical integral. It is our main task to
calculate $\mathcal C(\nu)$
in the case when $\nu$ is pure imaginary 
in Section~\ref{Sec:EFTI}.

As a warmup, we do the calculation of $\CC(\nu)$ for $0<\nu<3/2$
analytically in this
section. In this case, we first write
\be\label{A-B}
\CC(\nu)=\frac{\pi}{8}(\mathcal{A}-\mathcal{B}),
\ee
where $\mathcal{A}$ and $\mathcal{B}$ are the first and second terms, respectively,
of Eq.~(\ref{C(nu)}), and calculate the integrals
$\mathcal{A}$ and $\mathcal{B}$ piece by piece:
We use the asymptotic from of the Hankel function at $x\ll1$
for the integration from 0 to 1, and the asymptotic from at $x\gg1$
for the integration from 1 to $\infty$. We can see that,
the main contribution to the integral is from $0<x<1$, i.e.,
the infrared era when the wavelength
of the mode has been stretched out of horizon.

The asymptotic behavior of the Hankel function in the
ultraviolet limit is in the form of an ordinary wave,
but that in the infrared limit needs some caution
 since there is a spurious discontinuity and divergence at
integer $\nu$. To avoid it one should take more terms
of asymptotic expansion into account instead of
keeping only the leading order term.
 We defer the detailed discussion in Appendix~\ref{App:Hankel}.
Here we refer to the result given by Eqs.~(\ref{BesselJ def})
and (\ref{Neumann def}),
\bea
&&\mathcal{A} = \left|\int^1_0\dx~x^{-1/2}H_\nu^{(1)}(x)e^{ix}
  +\int^\infty_1\dx~x^{-1/2}H_\nu^{(1)}(x)e^{ix}\right|^2,
\nn
\\
&&\lra\left|\sum_{k=0}^\infty\frac{(-)^k}{2^{2k}k!}
\left\{\frac{1+i\cot\nu\pi}{2^\nu\Gamma(k+\nu+1)}\int^1_0\dx~
  x^{2k+\nu-1/2}e^{ix}\right.\right.
\nn\\
&&~-\left.\frac{i2^\nu}{\sin\nu\pi\Gamma(k-\nu+1)}
\int^1_0\dx~x^{2k-\nu-1/2}e^{ix}\right\}
  +\left.\int_1^\infty\dx~\sqrt{\frac{2}{\pi}}
\frac{e^{i(2x-\nu\pi/2-\pi/4)}}{x}\right|^2,
\nn\\
&&=\left|\sum_{k=0}^\infty\frac{(-)^k}{2^{2k}k!}
\left\{\frac{1+i\cot\nu\pi}{2^\nu\Gamma(k+\nu+1)}
  i^{2k+\nu+1/2}\gamma(2k+\nu+1/2,-i)\right.\right.
\nn\\
&&~-\left.\left.
\frac{i^{2k-\nu+3/2}2^\nu}{\sin\nu\pi\Gamma(k-\nu+1)}
\gamma(2k-\nu+1/2,-i)\right\}
  +\sqrt{\frac{2}{\pi}}e^{-i\frac{\pi}{2}\left(\nu+\frac{1}{2}\right)}
(\pi+i\mathrm E_1(2i))\right|^2,
\label{A final}
\eea
where some of the integrals are expressed in terms of
the incomplete gamma function,
\begin{equation}\label{def:gamma}
    \gamma(s,p)=\int^{p}_{0}t^{s-1}e^{-t}dt\,.
\end{equation}
The incomplete gamma function is defined originally on the real
$s$ axis, but can be analytically continued to the complex plane
with a branch cut from $-\infty$ to 0. 
 For a given wavelength, the third term in (\ref{A final}) is the
integral from the ultraviolet era $x>1$,
which contributes little to the final result.

The evaluation of $\mathcal{B}$ is a little tricky since there is an
overall time ordering operator. Keeping this in mind, we split
the integrals on two axes as
\begin{eqnarray}\label{B1}
  \mathcal{B}
&=& 2\re\int^1_0\dx_1~x_1^{-1/2}H_\nu^{(1)}(x_1)e^{-ix}
\int^1_{x_1}\dx_2~x_2^{-1/2}H_\nu^{(2)}(x_2)e^{-ix_2} \\
\label{B2}
&+& 2\re\int^1_0\dx_1~x_1^{-1/2}H_\nu^{(1)}(x_1)e^{-ix}
\int^\infty_1\dx_2~x_2^{-1/2}H_\nu^{(2)}(x_2)e^{-ix_2} \\
\label{B3}
&+& 2\re\int^\infty_1\dx_1~x_1^{-1/2}H_\nu^{(1)}(x_1)e^{-ix}
\int^\infty_{x_1}\dx_2~x_2^{-1/2}H_\nu^{(2)}(x_2)e^{-ix_2}\,.
\end{eqnarray}
As before, the second line (\ref{B2}) can be neglected since it
is small compared to the first line (\ref{B1}), while the
third line (\ref{B3}) can be discarded directly if one adds a
small imaginary part $i\epsilon$ to $x$ and do the integral
in the ultraviolet limit. Now the double integral in the
first line (\ref{B1}) is equivalent to the integration over
the upper triangle bounded by the $x_2$-axis, $x_2=1$ and $x_1=x_2$.
Since the integrand is in the form $2f(x_1)f(x_2)$, it is
in particular symmetric with respect to the interchange of
$x_1$ and $x_2$. Hence the integral is equal to a half of
that over the square bounded by $x_1=x_2=0$ and $x_1=x_2=1$.
The result is then given by the square of the integral
$\int_0^1dxf(x)$, that is,
\bea\nn
\mathcal{B}&\lra&\re\left[\int^1_0\dx~x^{1/2}Y_\nu(x)e^{-ix}\right]^2,
\\\nn
&=&\re\left[\sum_{k=0}^\infty\frac{(-)^k}{2^{2k}k!}
\left(\frac{\cot\nu\pi}{2^\nu\Gamma(k+\nu+1)}\int^1_0\dx~
  x^{2k+\nu-1/2}e^{-ix}\right.\right.
\\\nn
&&\left.\left.-\frac{2^\nu}{\sin\nu\pi\Gamma(k-\nu+1)}
\int^1_0\dx~x^{2k-\nu-1/2}e^{-ix}\right)\right]^2,
\\\nn
  &=&\re\left[\sum_{k=0}^\infty\frac{(-)^k}{2^{2k}k!}
\left(\frac{(-i)^{2k+\nu+1/2}\cot\nu\pi}{2^\nu\Gamma(k+\nu+1)}
  \Gamma(2k+\nu+1/2,0,i)\right.\right.\\
\label{B final}
&&\left.\left.-\frac{2^\nu(-i)^{2k-\nu+1/2}}{\sin\nu\pi\Gamma(k-\nu+1)}
\Gamma(2k-\nu+1/2,0,i)\right)\right]^2.
\eea

\begin{figure}[thb]
\includegraphics[width=0.6\textwidth]{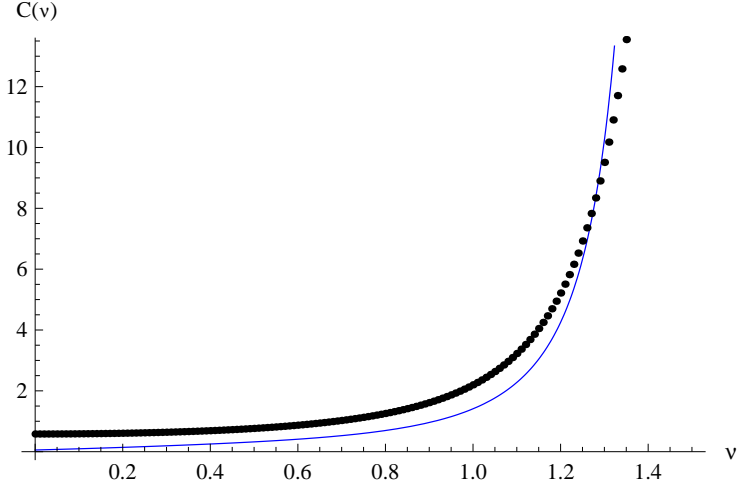}
\caption{The blue curve depicts an approximate analytic
expression for $\mathcal{C}(\nu)$ given by Eqs.~(\ref{A final}) and
(\ref{B final}), discarding the contributions coming from
the ultraviolet part $x>1$. The dots are those read from
the numerical result obtained in~\cite{Chen:2009zp}.
The summation over $k$ in these equations is truncated at $k\leq N=10^3$.}
\label{Fig:C(nu)}
\end{figure}

In summary, we obtain an approximate
analytical expression for $\mathcal{C}(\nu)$ by
combining (\ref{A-B}), (\ref{A final}) and (\ref{B final}).
It is depicted in Fig.~\ref{Fig:C(nu)} in which the sum
over $k$ is truncated at $N=10^3$.
We see that our approximate formula reproduces the one
obtained numerically in~\cite{Chen:2009zp} fairly well.
In fact, as discussed in detail in Appendix~\ref{App:Hankel},
the convergence of this series is very fast and
a truncation at $N=3$ is found to give a result
which is almost indistinguishable from that at $N=10^3$.


\section{Heavy Isocurvature Modes: Inflation under Effective Field Theory}
\label{Sec:EFTI}

\begin{figure}
  \includegraphics[width=0.6\textwidth]{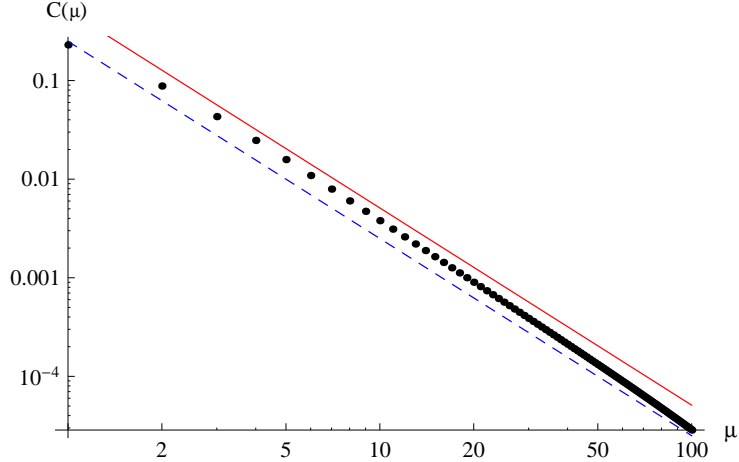}\\
  \caption{The red curve shows $\mathcal{C}(\mu)$ given by our analytical
approximation (\ref{CmuResult}), which is also the one derived from the effective
single-field approach~(\ref{cs in EFT}). The black dots are
the result of numerical integrations of (\ref{def:C(mu)}). There is a small deviation when $\mu\sim\mathcal{O}(10)$}
\label{Fig:Cmu}
\end{figure}

For an isocurvature mode which has a large effective mass, that is,
$\M/H>3/2$, the parameter $\nu$ becomes pure imaginary.
Therefore we introduce a real parameter $\mu$ by
\begin{equation}\label{mu}
    \nu\equiv i\mu=i\sqrt{\left(\frac{\M}{H}\right)^2-\frac{9}{4}}.
\end{equation}
Then the correction to the power spectrum is written as
\begin{eqnarray}\label{def:C(mu)}
 \mathcal{C}(\mu)&\equiv&\frac{\pi}{8}e^{-\mu\pi}(\alpha-\beta)\,;
\\
\alpha&=&
 \left|\int^\infty_0\dx~x^{-1/2}H_{i\mu}^{(1)}(x)e^{ix}\right|^2,
\\
\beta&=&
2\re\int^\infty_0\dx_1~x_1^{-1/2}H_{i\mu}^{(1)}(x_1)e^{-ix_1}
 \int^\infty_{x_1}\dx_2~x_2^{-1/2}(H_{i\mu}^{(1)}(x_2))^\ast e^{-ix_2}.
\end{eqnarray}
Note the existence of an overall exponential suppression factor
$e^{-\mu\pi}$. This kills almost all the contributions in
$\alpha$ and $\beta$ except for the one which contains an
exponential enhancement factor that exactly cancels it.
The above can be evaluated numerically as depicted in Fig.~\ref{Fig:Cmu}.
But we can evaluate it analytically
in the limit when the isocurvaton is very heavy, i.e. $\mu\gg1$,
which is the case when one expects an effective single-field
description holds.

First we consider $\alpha$. The integral in $\alpha$ is
\be\label{alphaintegral}
\int^\infty_0dx\;x^{1/2}H^{(1)}_{i\mu}(x)e^{ix}=\int^\infty_0dx\;x^{1/2}J_{i\mu}(x)e^{ix}+i\int^\infty_0dx\;x^{1/2}N_{i\mu}(x)e^{ix}.
\ee
The first term on the right hand side can be integrated directly as
\bea
\int^\infty_0dx\;x^{1/2}J_{i\mu}(x)e^{ix}&=&\frac{\Gamma(1/2+i\mu)}{2^{i\mu}(-i)^{1/2+i\mu}\Gamma(1+i\mu)}
{}_2F_1\left(\frac{1}{4}+\frac{i\mu}{2},\frac{3}{4}+\frac{i\mu}{2};1+i\mu;1\right),\\
&=&\frac{e^{i\pi/4}}{\sqrt{2\pi}}e^{-\mu\pi/2}.
\eea
In obtaining this we have used the special value of hypergeometric function when the last argument is 1, see (\ref{hypergeometric(1)}) in Appendix \ref{App:Gamma}, and relation \label{gamma+1/2} of gamma function. Next, we can calculate the second integral in (\ref{alphaintegral}).
There is a possible spurious divergence if we integrate it directly. Let us add a small positive imaginary part $i\epsilon$ to the exponent to make it converge, i.e. to write the exponent as $e^{ix-\epsilon x}$. Therefore, the integral is
\bea\nn
&&\int^\infty_0dx\;x^{1/2}N_{i\mu}(x)e^{ix}
\\\nn&=&\frac{\cot i\mu\pi\Gamma(1/2+i\mu)}{2^{i\mu}\Gamma(1+i\mu)(-2i\epsilon)^{1/4+i\mu/2}}\;\;
{}_2F_1\left(\frac{1}{4}+\frac{i\mu}{2},\frac{1}{4}+\frac{i\mu}{2};1+i\mu;\frac{1}{-2i\epsilon}\right)\\\label{alphaintegral2}
&-&\frac{2^{i\mu}\Gamma(1-i\mu)}{\sin i\mu\pi\Gamma(1/2-i\mu)(-2i\epsilon)^{1/4-i\mu/2}}\;\;
{}_2F_1\left(\frac{1}{4}-\frac{i\mu}{2},\frac{1}{4}-\frac{i\mu}{2};1-i\mu;\frac{1}{-2i\epsilon}\right).
\eea
Then this is the situation when the last argument of the Gaussian hypergeometric function is large, and (\ref{hypergeometric(z->1/z)}) can be applied to
connect it to
\bea\nn
(\ref{alphaintegral2})
&=&\frac{\dps\coth\mu\pi\Gamma\left(\half+i\mu\right)}{2^{i\mu}\Gamma(1+i\mu)}
\frac{\dps2\Gamma(1+i\mu)}{\dps\Gamma\left(\frac{1}{4}+\frac{i\mu}{2}\right)\Gamma\left(\frac{3}{4}+\frac{i\mu}{2}\right)}\;
{}_2F_1\left(\frac{1}{4}+\frac{i\mu}{2},\frac{1}{4}+\frac{i\mu}{2};1+i\mu;-2i\epsilon\right)\\\nn
&-&\frac{\dps2^{i\mu}\Gamma(\half-i\mu)}{\dps i\sinh\mu\pi\Gamma\left(1-i\mu\right)}
\frac{\dps 2\Gamma(\half-i\mu)}{\dps \Gamma\left(\frac{1}{4}-\frac{i\mu}{2}\right)\Gamma\left(\frac{3}{4}-\frac{i\mu}{2}\right)}\;
{}_2F_1\left(\frac{1}{4}-\frac{i\mu}{2},\frac{1}{4}-\frac{i\mu}{2};1-i\mu;-2i\epsilon\right).
\eea
Now the divergence disappears, and we can take $\epsilon\ra0$ and use (\ref{hypergeometric(0)}) to get
\be\label{alphaintegral3}
(\ref{alphaintegral2})
=\frac{\dps2\coth\mu\pi\Gamma\left(\half+i\mu\right)}{\dps 2^{i\mu}\Gamma\left(\frac{1}{4}+\frac{i\mu}{2}\right)\Gamma\left(\frac{3}{4}+\frac{i\mu}{2}\right)}\\\nn
-\frac{\dps 2^{1+i\mu}\Gamma\left(\dps\half-i\mu\right)}{\dps i\sinh\mu\pi\Gamma\left(\frac{1}{4}-\frac{i\mu}{2}\right)\Gamma\left(\frac{3}{4}-\frac{i\mu}{2}\right)}.
\ee
Next we take (\ref{gamma+1/2}) to cancel these gamma functions. We are left with
\be
(\ref{alphaintegral3})=\sqrt{\frac{2}{\pi}}\left(\coth\mu\pi+\frac{i}{\sinh\mu\pi}\right).
\ee
Therefore we have that
\be
\alpha=\frac{1}{\pi}\left|\frac{e^{\mu\pi/2}}{2}-\frac{\sqrt{2}}{\sinh\mu\pi}
+i\left(\frac{e^{-\mu\pi}}{2}+\sqrt{2}\coth\mu\pi\right)\right|^2.
\ee
This is an exact result, with no approximations. We see that when the effective isocurvature mass is large, this goes as
\be
\alpha\ra\frac{2}{\pi}\coth^2\mu\pi\sim1.
\ee
Together with the factor $e^{-\mu\pi}$ in (\ref{def:C(mu)}), we now that the contribution from $\alpha$ is exponentially suppressed.

Now we move to the evaluation of $\beta$,
\begin{equation}\label{beta}
 \beta=2\re\int^\infty_0\dx_1~x_1^{-1/2}H_{i\mu}^{(1)}(x_1)e^{-ix_1}
 \int^\infty_{x_1}\dx_2~x_2^{-1/2}(H_{i\mu}^{(1)}(x_2))^\ast e^{-ix_2}.
\end{equation}
Take the asymptotic behavior (\ref{Hankel1-mu approx}) we have got in Appendix \ref{App:HankelImaginary}, we can see
\bea\nn
\beta&=&\frac{4}{\pi}\frac{e^{\pi\mu}}{\mu}\re\int^\infty_0 dx_1x_1^{i\mu-1/2}\exp\left[-\frac{x_1^2}{4\mu}e^{-i\pi/4}-ix_1\right]\\\label{def:beta}
&&\times\int_{x_1}^\infty dx_2 x_2^{-i\mu-1/2}\exp\left[-\frac{x_2^2}{4\mu}e^{i\pi/4}-ix_2\right]
\eea
which contains a double integral. We first deal with the integral of $x_2$. As we have done in calculating $\alpha$, we can take the Taylor series of the exponent for $x_2$ and the integral of $x_2$ can be done term by term to get
\bea\label{InnerIntegral}&&
\int_{x_1}^{y_1} dx_2 x_2^{-i\mu-1/2}\exp\left[-\frac{x_2^2}{4\mu}e^{i\pi/4}-ix_2\right]\\\label{SumIntegral}
&=&\sk \int^{y_1}_{x_1}dx_2x_2^{-i\mu-1/2}\left(\frac{x_2^2}{4\mu}e^{i\pi/4}+ix_2\right)^k,\\\nn
&=&\sum_{k=0}^\infty\frac{(-)^k}{\dps k!\left(k+\frac{1}{2}-i\mu\right)} \frac{1}{4\mu}\\\nn
&\times&\left\{x_1^{k+1/2-i\mu}\left(i+\frac{(-)^{1/4}x_1}{4\mu}\right)^k[(-)^{3/4}x_1-4\mu]
{}_2F_1\left(1,\frac{3}{2}+2k-i\mu;\frac{3}{2}+k-i\mu;\frac{(-)^{3/4}x_1}{4\mu}\right)\right.\\\nn
&-&\left.y_1^{k+1/2-i\mu}\left(i+\frac{(-)^{1/4}y_1}{4\mu}\right)^k[(-)^{3/4}y_1-4\mu]
{}_2F_1\left(1,\frac{3}{2}+2k-i\mu;\frac{3}{2}+k-i\mu;\frac{(-)^{3/4}y_1}{4\mu}\right)\right\}\\\label{SingleIntegral}
\eea
Again, integrating $x_2$ from 0 to $\infty$ in (\ref{SumIntegral}) will give us a summation of infinities, which is unphysical since
we know that in (\ref{InnerIntegral}) there is an exponential suppression which will converge the integral. Thus, we take a UV cutoff at $y_1$, and set $y_1\ra\infty$ after all the integrations and summations are done. Thus this integral can be expressed by its lower bound $x_1$ and upper bound $y_1$ as above via the hypergeometric function ${}_2F_1$. We make use of the asymptotic behavior of this function when $\mu\gg1$ in Appendix $\ref{App:Gamma}$.
From (\ref{hypergeometric(small)}), we see that when $\mu$ is large and $k\ll\mu$, the hypergeometric functions with different $k$'s are the same, and we can resum the polynomials to get back to an exponent, which yields
\be\label{SingleIntegralResult}
(\ref{SingleIntegral})=\frac{i}{\mu}\left\{y_1^{1/2-i\mu}\exp\left[-y_1\left(i+\frac{(-)^{1/4}y_1}{4\mu}\right)\right]
-x_1^{1/2-i\mu}\exp\left[-x_1\left(i+\frac{(-)^{1/4}x_1}{4\mu}\right)\right]\right\}.
\ee
Now we can take the limit of integration upper bound $y_1$ to be infinite.
This kills the first term in (\ref{SingleIntegralResult}). Then,
substituting (\ref{SingleIntegralResult}) into (\ref{def:beta}),
\be\nn
\beta=-\frac{4}{\pi}
\frac{e^{\pi\mu}}{\mu^2}\re\;i\int^\infty_0dx_1
\exp\left[-\frac{\sqrt{2}x_1^2}{4\mu}-2ix_1\right].
\ee
This integral can be expressed by the error function,
\be\label{def:efi}
\mathrm{erf}(z)=\frac{2}{\sqrt{\pi}}\int_0^z e^{-t^2}dt\,,
\ee
as
\be
\beta=2^{7/4}\frac{e^{\mu\pi}}{\sqrt{\pi\mu^3}}e^{-2\sqrt{2}\mu} \mathrm{Re}[i\mathrm{erf}(i2^{3/4}\mu)].
\ee
We can see that it depends on a power-law series of $\mu$ when $\mu$ is large. To be specific, let us can take the Taylor series of $\beta$ around infinity to get
\be\label{betaResult}
\beta=-2\frac{e^{\mu\pi}}{\pi\mu^2}\left[1+\sum_{k=1}^\infty\frac{(2k-1)!!}{2^{5k/2}\mu^k}\right].
\ee
This is the only term that contributes to the final correction to
the power spectrum, and we see that the leading term is proportional to
 $\mu^{-2}$(thus to $\M^2$) when the effective mass of the isocurvature mode is
large. We take this leading term and go back from the combination of (\ref{P-C}),
 (\ref{def:C(mu)}) and (\ref{betaResult}) to obtain our final result
\bea\label{CmuResult}
\mathcal C(\mu)&\approx&\frac{1}{4\mu^2},
\\\label{SpectrumResult}
\mathcal P_{\cal R}&\approx&\mathcal{P}_{\cal R}^{(0)}\left[1+2\frac{H^2}{M_\mathrm{eff}^2}
\left(\frac{\dot\theta}{H}\right)^2\right]\,.
\eea
As before $\approx$ denotes neglecting the terms of higher power.
Note the dependence on $\mu$. It is not an exponential
suppression, but a power-law one.
We reached (\ref{SpectrumResult}) under the approximation
that $\mu$ is large as well as $\dot\theta/H$ is small.
As depicted in Fig.~\ref{Fig:Cmu},
the analytic approximation (\ref{CmuResult})
reproduces a numerical evaluation of the formula~(\ref{def:C(mu)})
well. This is our main result.

To compare this with the one obtained by the
effective-single-field approach, let us refer to the result
in~\cite{Tolley:2009fg,Achucarro:2010da,Cespedes:2012hu,Achucarro:2012sm}
where the two-field case with large isocurvature mass is considered.
After integrating out the heavy field, the power spectrum is
corrected in terms of the effective sound speed as
\begin{equation}\label{cs in EFT}
c_s^{-2}=1+\frac{4H^2}{\tilde M_\mathrm{eff}^2}
\left(\frac{\dot\theta}{H}\right)^2,
\end{equation}
where to leading order
$\tilde M_\mathrm{eff}^2=V_{\sigma\sigma}-\dot\theta^2$
in our notation.
We first note that the definition of effective mass
$\tilde M_\mathrm{eff}^2$ is different from ours
by $4\dot\theta^2$, which is
not important since
 it is assumed to be small when the perturbative in-in formulism can be applied.
 Besides, we also see that our result (\ref{CmuResult}), which gives
the power spectrum
a correction factor corresponding
to $c_s^{-1}$, is exactly the same as (\ref{cs in EFT}) at leading order
when $\dot\theta^2$ is small. So we have shown the equivalence of both methods applying on the same model.

\section{Conclusions and Discussions}
\label{Sec:conclude}
In this paper, we studied a two-field model of inflation in which
the curvature mode, that is the mode along the inflationary trajectory,
is light while the isocurvature mode, that is perpendicular to the trajectory, is heavy. To simplify the problem,
we focused on a situation in which the field is in a circular motion
with a small constant angular velocity.
By computing a process that describes the effect of the heavy
isocurvature mode on the curvature perturbation 2-point function by the
in-in formalism, we derived a correction to the curvature perturbation power
spectrum due to the heavy mode intermediation. The result tells us that the correction is proportional to $(\dot\theta/\M)^2$,
which is exactly the same as
the one obtained by the effective single-field approach by integrating out the heavy field.

Here let us reconsider our assumptions, and discuss the validity
and limitaions of both methods.
We already know that only if
\be\label{slow turn}
\frac{\dot\theta^2}{H^2}\ll1\,,
\ee
one can use perturbation theory in the in-in formalism which we employed.
On the other hand, in the case of a sharp turn with large angular velocity, even for a very short time,
our method fails.
 It is demonstrated in~\cite{Shiu:2011qw,Gao:2012uq} that such a case cannot be
described by an effective single-field action either. The other assumption is
$M_\mathrm{eff}^2\gg H^2$ which we use to get our final analytic expressions. Although this is not
necessary for the use of the in-in formulism, it
enabled us to obtain a simple analytic result and to easily compare this with the effective single-field approach
in which the heavy modes are integrated out.

Some papers discussed the requirements of effective single field theory
 to hold besides the isocurvature mode is heavy. The main constraint is
 called the adiabatic condition~\cite{Cespedes:2012hu},
\be
\left|\frac{\text d}{\text dt}\log\dot\theta\right|\ll M_\text{eff}\,,
\quad\text{or}\quad
\left|\frac{\text d}{\text dt}\log(c_s^{-2}-1)\right|\ll M_\text{eff}\,.
\ee
In our constant turn case these conditions are satisfied
since we have $\ddot\theta=0$. It is shown that the decoupling of the heavy field by integrating it out is valid only when slow-roll condition is preserved\cite{Avgoustidis:2012yc}. In~\cite{Shiu:2011qw} the authors claim that to keep slow-roll in the classical trajectory one should
impose
\be
\frac{V''}{H}\gg\frac{\dot\theta}{H}\,,
\ee
which is satisfied for $\mu\gg1$.
On the other hand, for $\nu\ra3/2$ which corresponds to the case
when not only the inflaton but also the isocurvaton
have their masses negligible, there does not exist
even an appropriate classical quasi-single-field trajectory.
This is probably the reason why we see the divergent behavior
in Fig.~\ref{Fig:C(nu)} at $\nu\ra3/2$.
In this case, one should abandon the effective
single-field description, and turn to other methods
used in the standard slow-roll multi-field case as
mentioned in Section \ref{Sec:intro}.

It is important to study the implications of
quasi-single field models in realistic situations.
It surely represents a kind of turning process, but it must be adiabatic.
If the trajectory is straight before and after the adiabatic turning,
the power spectrum (\ref{SpectrumResult}) is only modified during
the turning process, which implies a feature at a characteristic
scale determined by the time-dependence of $\dot\theta$.
It will be interesting to study this ``adiabatic turn'' quasi-single field
inflation in more detail. For instance, we know this scale-dependence can only be detected if the
turning occurs not too late at the inflationary stage.
In hybrid inflation~\cite{Linde:1993cn}, inflation ends by a sharp
 turn in the field space and the mode that was isocurvature during
inflation turns into the waterfall field and becomes
important~\cite{Lyth:2010ch,Lyth:2010zq,Fonseca:2010nk,Abolhasani:2010kn,Gong:2010zf}.
In particular, recently it was pointed out that a spiky feature may
appear in the spectrum and bispectrum for a certain parameter range
of hybrid inflation~\cite{Abolhasani:2012px}.
The waterfall process is not adiabatic. Thus our method cannot be applied.
Nevertheless, it is interesting to see if an extension of the present
method based on the in-in formalism can shed more light on these
cases of hybrid inflation.

\begin{figure}
  \begin{minipage}[t]{0.35\linewidth}
    \centering
    \includegraphics[width=0.8\textwidth]{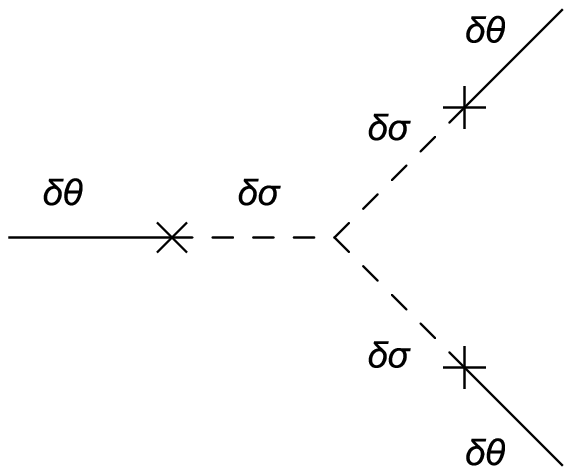}
  \caption{This is the main contribution to large non-Gaussianity evaluated in~\cite{Chen:2009zp}.}
    \label{Fig:3p3}
  \end{minipage}%
  ~~~~~~~~
  \begin{minipage}[t]{0.66\linewidth}
      \begin{minipage}[t]{0.5\linewidth}
    \centering
    \includegraphics[width=0.8\textwidth]{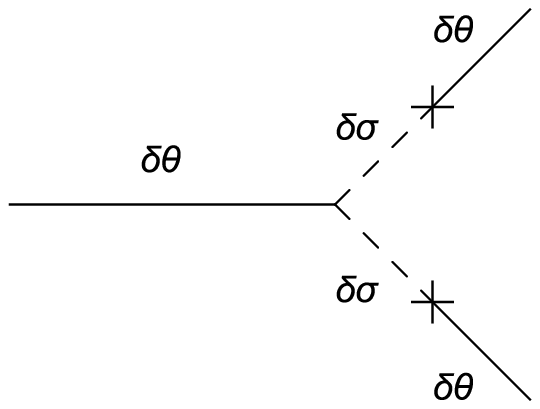}
    \end{minipage}~~
    \begin{minipage}[t]{0.5\linewidth}
    \centering
    \includegraphics[width=0.8\textwidth]{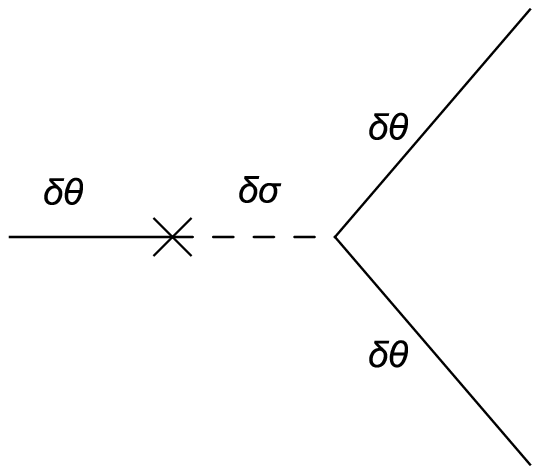}
  \end{minipage}
      \caption{These are other possible subleading
      contributions to the non-Gaussianity in quasi-single field inflation.}
    \label{Fig:3p1&2}
  \end{minipage}
\end{figure}

Finally, let us comment on possible developments from our current work.
 We notice that our result (\ref{SpectrumResult})
 is exact only at leading order in the limit  $\mu\ra\infty$.
 Beyond this limit we can calculate higher order correction in $1/\M$ to compensate
possible deviations from the leading order result.
An interesting possibility emerges when we consider next-to-leading order terms.
We know in the EFT approach, corrections to the power spectrum from the
isocurvaton come in only via $V_{\sigma\sigma}$, which means
the next-to-leading order correction should be of order
$1/V_{\sigma\sigma}^2\sim1/\tilde M_\mathrm{eff}^4$.
On the other hand, in our approach, the effect comes from
$\mu=\sqrt{\M^2/H^2-9/4}\sim\M$, thus naively the next-to-leading correction
will be $1/\M^3$. Therefore, unless we has a mechanism to cancel all the contributions
from the terms with odd power indices,
we will face this inconsistency. We need further investigations to
find out the possible implications behind this fact.

Another access to check this consistency is to consider the higher order correlations, specifically
the 3-point function of the curvature perturbation.
Actually, one of the main motivations for developing
the quasi-single field approach was to calculate the non-Gaussianity
expected in such a model~\cite{Chen:2009zp,Assassi:2012zq}.
It is found that only the vertex given by $V'''$, i.e.,
the last term in (\ref{CH3}) corresponding to the diagram shown in
Fig.~\ref{Fig:3p3} was of most importance in contributing to large non-Gaussianities.
This is because the self-interaction of the isocurvature mode, i.e. $V'''$,
is possible to be large, while the other vertices in (\ref{CH3}), as are depicted in
Fig.~\ref{Fig:3p1&2}, will give a negligible contribution. Under the same assumptions,
 we can also calculate
the similar coefficient $s(\nu)$ and $\alpha(\nu)$ of non-Gaussian parameter $f_\mathrm{NL}$ in the three-point correlations
which also depend on the effective mass,
extrapolate it to the case with a large isocurvaton mass, and compare it with
the non-Gaussianity in effective field approach~\cite{Tolley:2009fg,Achucarro:2010da}.
It is very interesting to see whether these two results still coincides with
each other, and if so, whether we can prove the equivalence of the two approaches
to all orders.
\\

\textbf{Note}: When this work was nearly completed, we were acquainted by
 private communication that a related work~\cite{Chen:2012ge} on this issue was
also approaching its completion. We thank the authors of \cite{Chen:2012ge}
for synchronizing our submission to the arXiv.

\begin{acknowledgements}
We are grateful to Yifu Cai for useful discussions on numerical
methods. SP thanks the hospitality during his visit to Yukawa Institute
for Theoretical Physics for the workshop ``2012 Asia Pacific School/Workshop on Cosmology and Gravitation"
(YITP-W-11-26) when this work was
initiated. SP also thanks Bin Chen, Qiang Xu, Jiaju Zhang for
 valuable comments on the work. We thank the referee and editor of JCAP, who have pointed out the inconsistency between our result in the first version and EFT's, and give us suggestions on how to improve the accuracy in our calculation.
This work is supported in part by the JSPS Grant-in-Aid for
Scientific Research (A) No.~21244033, and by the
MEXT Grant-in-Aid for the global COE program at Kyoto University,
``The Next Generation of Physics, Spun from Universality and Emergence".
SP is supported by the NSFC Grant No.10975005 and Scholarship
Award for Excellent Doctoral Student granted by Ministry of
Education of China.
\end{acknowledgements}

\appendix
\section{The Asymptotic Form of Hankel Functions of Real Order}
\label{App:Hankel}

Here we give several useful expressions for the asymptotic expansion
of the Hankel functions, paying particular attention to the case of
integer orders

We start from the series definition of the Bessel function,
\bea\label{BesselJ def}
J_\nu(x)\lra\sum^\infty_{k=0}\frac{(-)^k}{k!}
\frac{1}{\Gamma(k+\nu+1)}\hox^{2k+\nu}\,.
\eea
From this definition the Neumann function is defined by
\bea
\label{Neumann def}
Y_\nu(x)
&=&\cot\nu\pi J_\nu(x)-\frac{1}{\sin\nu\pi}J_{-\nu}(x)
\nn\\
&=&\sum^\infty_{k=0}\frac{(-)^k}{k!}\hox^{2k}
\left[\frac{\cot\nu\pi}{\Gamma(k+\nu+1)}\hox^\nu
-\frac{1}{\sin\nu\pi\Gamma(k-\nu+1)}\hox^{-\nu}\right].
\eea
For$\nu>0$, it seems that we can neglect the first term
in (\ref{Neumann def}) since it is small compared to the second one
when $x\ra0$. However, both terms suffer from divergence
when $\nu$ is an integer, hence the first term cannot be simply discarded.

To see how the divergence in each term is canceled with each other,
let us consider first the behavior at $\nu=0$.
In this limit, we have
\bea\nn
\lim_{\nu\ra0}Y_\nu(x)
&=&\frac{1}{\nu\pi}\left[1+\nu\ln\hox\right]
-\frac{1}{\nu\pi}\left[1-\nu\ln\hox\right]\\\nn
&&+\sum^\infty_{k=1}\frac{(-)^k}{k!}
\left\{\frac{1}{\nu\pi\Gamma(k+1)}\left[1+\nu\ln\hox\right]
-\frac{1}{\nu\pi\Gamma(k+1)}\left[1-\nu\ln\hox\right]\right\},\\
&=&\frac{2}{\pi}J_0(x)\ln\frac{x}{2}\\
&\lra&\frac{2}{\pi}\ln\frac{x}{2}\quad\mathrm{for}~~x\ra0.
\eea
Thus we obtain a finite result because of the cancelation of $1/\nu$
divergences coming from both terms.

Now consider the case when $\nu$ is an integer.
In our case, since $\nu\leq3/2$, only the case $\nu=1$ concerns us.
Introducing $\epsilon\equiv1-\nu$, we have
\bea\nn
\lim_{\nu\ra1}Y_\nu(x)&=&\sum^\infty_{k=1}
\frac{(-)^k}{k!}\hox^{2k}\left[
-\frac{1}{\ep\pi\Gamma(k+2)}\hox\left(1-\ep\ln\hox\right)\right.
\nn\\
&&\left.-\frac{1}{\ep\pi\Gamma(k+\ep)}
\hox^{-1}\left(1+\ep\ln\hox\right)\right]
\nn\\
&=&-\frac{2}{\pi x}+\frac{2}{\pi}J_1(x)\ln\hox,
\nn\\
&\lra&-\frac{2}{\pi x}~~\mathrm{when}~~x\ra0.
\eea
Thus there is no divergence as $1/(1-\nu)$.

Something subtle appears when we want to truncate the
summation~(\ref{Neumann def}) to finite terms. From the calculation above,
we see that each term in the series is divergent,
although the residue is suppressed exponentially
by $1/\Gamma(k+1)\Gamma(k+\nu+1)$ or $1/\Gamma(k+1)\Gamma(k-\nu+1)$
 when $k$ is large.
Expanding~(\ref{Neumann def}), we find that for an integer $\nu$,
the divergence of the $k^+$-th term of positive series, $x^{2k^++\nu}$,
 is canceled by the divergence of the $k^-$-th term of the negative series,
 $x^{2k^--\nu}$, where $k^-=k^++\nu$. Truncating the $k^+$ and $k^-$ series
simultaneously at the same integer $N$ will cancel the divergence at $\nu=0$,
 but all the other divergences at $\nu\ge1$ survive. Similarly, one can also
truncated at $k^+=N$ and $k^-=N+m$, where $m$ is a positive integer
to cancel the divergence at $\nu=m$, but divergences at all the other
integers remain. Thus unless we take the infinite series limit $N\ra\infty$,
 all the divergences at all the integers cannot be canceled simultaneously.

Fortunately, however, the residues of these divergences are
 suppressed by $\Gamma(N)$ for large $N$. So in practical contexts
this will not cause a problem. For example, if we truncate the series
at $k^+=k^-=1000$ to cancel the divergence at $\nu=0$, the one at $\nu=1$
 remains. However since the residue is suppressed by $1/\Gamma(1000)^2$,
the divergence becomes effectively invisible unless we take the value
of $\nu$ exponentially close to $\nu=1$, within the width of
$\Delta\nu\sim1/\Gamma(1000)^2$. This enables us to draw
our analytic curve practically smooth as depicted
in Fig.~\ref{Fig:C(nu)}.

With the above understanding, we obtain the Hankel functions
in the form,
\begin{eqnarray}\label{def:H1}
  H^{(1)}(x) &=& J_\nu(x)+iY_\nu(x)\rightarrow iY_\nu(x);
\\\label{def:H2}
  H^{(2)}(x) &=& J_\nu(x)-iY_\nu(x)\rightarrow -iY_\nu(x)
\end{eqnarray}
where the Neumann function is given by the form~(\ref{Neumann def}).
The last arrows above hold only for a real $\nu$ but not for
an imaginary $\nu$, $\nu=i\mu$ for $\mu>0$.

As for the ultraviolet behavior at $x\gg1$, it is trivial,
\bea\label{Hankel1-nu UV}
H_{\nu}^{(1)}&\approx&\sqrt{\frac{2}{\pi x}}e^{i(x-\nu\pi/2-\pi/4)},
\\\label{Hankel2-nu UV}
H_{\nu}^{(2)}&\approx&\sqrt{\frac{2}{\pi x}}e^{-i(x-\nu\pi/2-\pi/4)},
\\\label{Hankel1-mu UV}
H_{i\mu}^{(1)}&\approx&\sqrt{\frac{2}{\pi x}}e^{\mu\pi/2}e^{i(x-\pi/4)},
\\\label{Hankel2-mu UV}
H_{i\mu}^{(2)}&\approx&\sqrt{\frac{2}{\pi x}}e^{-\mu\pi/2}e^{-i(x-\pi/4)}.
\eea
These formulae hold for any complex order $\nu$.

\section{Gamma Function and Hypergeometric Function}
\label{App:Gamma}

A useful relation which is used in the context is the product of two gamma functions whose arguments vary by 1/2:
\be\label{gamma+1/2}
\Gamma(z)\Gamma\left(\frac{1}{2}+z\right)=\frac{2\sqrt{\pi}}{2^{2z}}\Gamma(2z).
\ee

Next let us refer to the asymptotic expansion of
the Gamma function $\Gamma(z)$ at $z\to\infty$ on complex plane,
\begin{equation}\label{Gamma asymp}
    \Gamma(z)\approx(z)^{z-1/2}e^{-z}\sqrt{2\pi}
\left\{1+\frac{1}{12z}-\frac{1}{288z^2}+\dots\right\}.
\end{equation}
In this paper the most commonly used argument is like $a+i\mu$ where $\mu\gg1$ and $a\ll\mu$. Substituting it into (\ref{Gamma asymp}) we have
\be\label{gamma approx}
\Gamma(a\pm i\mu)\approx
\sqrt{2\pi}e^{\pm i[\mu\ln\mu-\mu+\pi(a-1)/4]}\mu^{a-1/2}e^{\pi\mu/2}.
\ee
Here for simplicity we only preserve the leading term.

The Gaussian hypergeometric function ${}_2F_1(a,b;c;z)$ appears when dealing with the integrals
in section~\ref{Sec:EFTI},
\be\label{def:hypergeometric}
{}_2F_1(a,b;c;z)\equiv\sum_{n=0}^\infty\frac{\Gamma(a+n)\Gamma(b+n)\Gamma(c)}{\Gamma(a)\Gamma(b)\Gamma(c+n)}\frac{z^n}{n!}.
\ee
We will use the hypergeometric function in three different limit of $z$. First, from the definition, we have
\be\label{hypergeometric(1)}
{}_2F_1(\alpha,\beta;\gamma;1)=\frac{\Gamma(\gamma)\Gamma(\gamma-\alpha-\beta)}{\Gamma(\gamma-\alpha)\Gamma(\gamma-\beta)}.
\ee

Now deal with large $z$, which can be connected with the hypergeometric function with a small argument $1/z$ as
\bea\label{hypergeometric(z->1/z)}
\frac{\Gamma(\alpha)\Gamma(\beta)}{\Gamma(\gamma)}{}_2F_1(\alpha,\beta;\gamma;z)
&=&\frac{\Gamma(\alpha)\Gamma(\beta-\alpha)}{\Gamma(\gamma-\alpha)}(-z)^{-\alpha}{}_2F_1\left(\alpha,\alpha-\gamma+1;\alpha-\beta+1;\frac{1}{z}\right)\\
&+&\frac{\Gamma(\beta)\Gamma(\alpha-\beta)}{\Gamma(\gamma-\beta)}(-z)^{-\beta}{}_2F_1\left(\beta,\beta-\gamma+1;\beta-\alpha+1;\frac{1}{z}\right).
\eea
It holds when $|z|>1$ and $z$ is not a positive real number.

The other limit is $z\ra0$. In this paper we are interested in the arguments $a=1$, $b=3/2+2k-i\mu$ and $c=3/2+k-i\mu$. Thus we have
\be
{}_2F_1\left(1,\frac{3}{2}+2k-i\mu;\frac{3}{2}+k-i\mu;z\right)
=\sum_{n=0}^\infty\frac{\dps\Gamma\left(\frac{3}{2}+2k-i\mu+n\right)\Gamma\left(\frac{3}{2}+k-i\mu\right)}
{\dps\Gamma\left(\frac{3}{2}+2k-i\mu\right)\Gamma\left(\frac{3}{2}+k-i\mu+n\right)}z^n.
\ee
When we take the limit $\mu\ra\infty$, $k\ll\mu$ and $n\ll\mu$,
using (\ref{gamma approx}), we can resum the power series to get
\be\label{hypergeometric(small)}
{}_2F_1\left(1,\frac{3}{2}+2k-i\mu;\frac{3}{2}+k-i\mu;z\right)
\approx\sum_{n=0}^\infty z^n=\frac{1}{1-z}\,.
\ee
The approximation holds because $|z|<1$ and the main contribution
 comes from the terms with small $n$ ($\ll\mu$),
which makes it possible to push the summation upper bound to infinity.
When we take $z\ra0$, we have
\be\label{hypergeometric(0)}
{}_2F_1(\alpha,\beta;\gamma;0)=1
\ee
which can also obtained by setting $z=0$ in the definition (\ref{def:hypergeometric}).


\section{Hankel Function of Imaginary Order}\label{App:HankelImaginary}
In this appendix we turn to the Hankel function of imaginary order $i\mu$,
focusing on the large $\mu$ limit,
which corresponds to the isocurvaton with a large mass.
For an imaginary order, in contrast to the case discussed in
 Appendix~\ref{App:Hankel}, we cannot neglect $J_{i\mu}$.
We have
\bea
H_{i\mu}^{(1)}(x)
&=&\sum^\infty_{k=0}\frac{(-)^k}{k!}\frac{1+i\cot i\mu\pi}{\Gamma(k+1+i\mu)}
\left(\frac{x}{2}\right)^{2k+i\mu}
\nonumber\\
&&-\frac{i}{\sin i\mu\pi}\sum^\infty_{k=0}\frac{(-)^k}{k!}
\frac{1}{\Gamma(k-1+i\mu)}\left(\frac{x}{2}\right)^{2k-i\mu}.
\label{Hankel1-mu}
\eea
We consider the large mass limit of (\ref{Hankel1-mu}), that is,
$\mu\rightarrow\infty$.
First note that in this limit, we have
\bea
\cot i\mu\pi=-i\coth\mu\pi&\ra&-i,\\
\sin i\mu\pi=i\sinh\mu\pi&\ra&\frac{ie^{\mu\pi}}{2}\,.
\eea
The above expressions imply that the second line of (\ref{Hankel1-mu})
is exponentially suppressed.
Besides, since the large $k$ terms in the summation of the first line of
(\ref{Hankel1-mu}) are also highly suppressed by the factor
 $(\Gamma(k+1)\Gamma(k+1-i\mu))^{-1}$, we expect the main contribution
to this summation comes from the terms with small $k$.
This permits us to apply $k\ll\mu$ in the Gamma function,
and use the approximate form given in (\ref{gamma approx}),
\be
\Gamma(k+1+i\mu)\approx\sqrt{2\pi}e^{i(\mu\ln\mu-\mu+k\pi/4)}
\mu^{k+1/2}e^{-\mu\pi/2}\,.
\ee
Therefore, we can pick out the $k$-dependence in the Gamma function,
and resum the polynomial to get
\be\label{Hankel1-mu approx}
H_{i\mu}^{(1)}\approx e^{-i\mu(\ln\mu-1)}\sqrt{\frac{2e^{\pi\mu}}{\pi\mu}}\exp\left[-\frac{x^2}{4\mu}e^{-i\frac{\pi}{4}}\right]
\left(\frac{x}{2}\right)^{i\mu}.
\ee



\begin{thebibliography}{99}


\bibitem{inflation_bible}
  A.~H.~Guth,
  ``The Inflationary Universe: A Possible Solution To The Horizon And Flatness
  Problems,''
  Phys.\ Rev.\  D {\bf 23}, 347 (1981);
\\
  K.~Sato,
 ``First Order Phase Transition Of A Vacuum And Expansion Of The Universe,''
  Mon.\ Not.\ Roy.\ Astron.\ Soc.\  {\bf 195}, 467 (1981).

\bibitem{newinflation}
   A.~A.~Starobinsky,
  ``A new type of isotropic cosmological models without singularity,''
  Phys.\ Lett.\  B {\bf 91}, 99 (1980);
\\
  A.~D.~Linde,
  ``A New Inflationary Universe Scenario: A Possible Solution Of The Horizon,
  Flatness, Homogeneity, Isotropy And Primordial Monopole Problems,''
  Phys.\ Lett.\  B {\bf 108}, 389 (1982);
\\
  A.~Albrecht and P.~J.~Steinhardt,
  ``Cosmology For Grand Unified Theories With Radiatively Induced Symmetry
  Breaking,''
  Phys.\ Rev.\ Lett.\  {\bf 48}, 1220 (1982).


\bibitem{Komatsu:2010fb}
  E.~Komatsu {\it et al.},
  ``Seven-Year Wilkinson Microwave Anisotropy Probe (WMAP) Observations:
  Cosmological Interpretation,''
  arXiv:1001.4538 [astro-ph.CO].


\bibitem{Susskind:2003kw}
  L.~Susskind,
  ``The Anthropic landscape of string theory,''
  In *Carr, Bernard (ed.): Universe or multiverse?* 247-266
  [hep-th/0302219].

\bibitem{inin}
J. S. Schwinger, ``The Special Canonical Group'', Proc. Nat. Acad. Sci. 46, 1401 (1960);~
  P.~M.~Bakshi and K.~T.~Mahanthappa,
  ``Expectation value formalism in quantum field theory. 1,''
  J.\ Math.\ Phys.\  {\bf 4}, 1 (1963);~
  P.~M.~Bakshi and K.~T.~Mahanthappa,
  ``Expectation value formalism in quantum field theory. 2,''
  J.\ Math.\ Phys.\  {\bf 4}, 12 (1963);~
L. V. Keldysh, Zh. Eksp. Teor. Fiz. 47, 1515 (1964).


\bibitem{Weinberg:2005vy}
  S.~Weinberg,
  ``Quantum contributions to cosmological correlations,''
  Phys.\ Rev.\  D {\bf 72}, 043514 (2005).
  [arXiv:hep-th/0506236].
\bibitem{Maldacena:2002vr}
  J.~M.~Maldacena,
  ``Non-Gaussian features of primordial fluctuations in single field
  inflationary models,''
  JHEP {\bf 0305}, 013 (2003)
  [arXiv:astro-ph/0210603].

\bibitem{Bunch:1978yq}
  T.~S.~Bunch and P.~C.~W.~Davies,
  ``Quantum Field Theory In De Sitter Space: Renormalization By Point
  Splitting,''
  Proc.\ Roy.\ Soc.\ Lond.\  A {\bf 360}, 117 (1978).

\bibitem{Liddle:1998jc}
  A.~R.~Liddle, A.~Mazumdar and F.~E.~Schunck,Dimopoulos:2005ac
  ``Assisted inflation,''
  Phys.\ Rev.\  D {\bf 58}, 061301 (1998)
  [arXiv:astro-ph/9804177].

\bibitem{Silverstein:2003hf}
  E.~Silverstein and D.~Tong,
  ``Scalar speed limits and cosmology: Acceleration from D-cceleration,''
  Phys.\ Rev.\  D {\bf 70}, 103505 (2004)
  [arXiv:hep-th/0310221];



\bibitem{Huang:2007hh}
  M.~X.~Huang, G.~Shiu and B.~Underwood,
  ``Multifield DBI Inflation and Non-Gaussianities,''
  Phys.\ Rev.\  D {\bf 77}, 023511 (2008)
  [arXiv:0709.3299 [hep-th]].
\bibitem{Ward:2007gs}
  J.~Ward,
  ``DBI N-flation,''
  JHEP {\bf 0712}, 045 (2007)
  [arXiv:0711.0760 [hep-th]].

\bibitem{Pi:2011tv}
  S.~Pi and D.~Wang,
  ``Cosmological perturbations in inflation with multiple sound speeds,''  Nucl.\ Phys.\ B {\bf 862}, 409 (2012)  [arXiv:1107.0813 [hep-th]].



\bibitem{Sasaki:1998ug}
  M.~Sasaki and T.~Tanaka,
  ``Superhorizon scale dynamics of multiscalar inflation,''
  Prog.\ Theor.\ Phys.\  {\bf 99}, 763 (1998)
  [gr-qc/9801017].

\bibitem{Gordon:2000hv}
  C.~Gordon, D.~Wands, B.~A.~Bassett and R.~Maartens,
  ``Adiabatic and entropy perturbations from inflation,''
  Phys.\ Rev.\  D {\bf 63}, 023506 (2001)
  [arXiv:astro-ph/0009131].

\bibitem{Amendola:2001ni}
  L.~Amendola, C.~Gordon, D.~Wands and M.~Sasaki,
  ``Correlated perturbations from inflation and the cosmic microwave
  background,''
  Phys.\ Rev.\ Lett.\  {\bf 88}, 211302 (2002)
  [arXiv:astro-ph/0107089].

\bibitem{Peterson:2010np}
  C.~M.~Peterson and M.~Tegmark,
  ``Testing Two-Field Inflation,''
  Phys.\ Rev.\  D {\bf 83}, 023522 (2011)
  [arXiv:1005.4056 [astro-ph.CO]].


\bibitem{ArmendarizPicon:1999rj}
  C.~Armendariz-Picon, T.~Damour and V.~F.~Mukhanov,
  ``k-Inflation,''
  Phys.\ Lett.\  B {\bf 458}, 209 (1999)
  [arXiv:hep-th/9904075].



\bibitem{Gao:2009qy}
  X.~Gao,
  ``On Cross-correlations between Curvature and Isocurvature Perturbations
  during Inflation,''
  JCAP {\bf 1002}, 019 (2010)
  [arXiv:0908.4035 [hep-th]].

\bibitem{Mukhanov:1982nu}
  V.~F.~Mukhanov and G.~V.~Chibisov,
  ``The Vacuum energy and large scale structure of the universe,''
  Sov.\ Phys.\ JETP {\bf 56}, 258 (1982)
  [Zh.\ Eksp.\ Teor.\ Fiz.\  {\bf 83}, 475 (1982)].

\bibitem{Sasaki:1986hm}
  M.~Sasaki,
  ``Large Scale Quantum Fluctuations in the Inflationary Universe,''
  Prog.\ Theor.\ Phys.\  {\bf 76}, 1036 (1986).

\bibitem{Mukhanov:1988jd}
  V.~F.~Mukhanov,
  ``Quantum Theory of Gauge Invariant Cosmological Perturbations,''
  Sov.\ Phys.\ JETP {\bf 67}, 1297 (1988)
  [Zh.\ Eksp.\ Teor.\ Fiz.\  {\bf 94N7}, 1 (1988)].

\bibitem{Sasaki:1995aw}
  M.~Sasaki and E.~D.~Stewart,
  ``A General Analytic Formula For The Spectral Index Of The Density
  Perturbations Produced During Inflation,''
  Prog.\ Theor.\ Phys.\  {\bf 95}, 71 (1996)
  [arXiv:astro-ph/9507001].
\bibitem{Lyth:2004gb}
  D.~H.~Lyth, K.~A.~Malik and M.~Sasaki,
  ``A general proof of the conservation of the curvature perturbation,''
  JCAP {\bf 0505}, 004 (2005)
  [arXiv:astro-ph/0411220].

\bibitem{Starobinsky:1986fxa}
  A.~A.~Starobinsky,
  ``Multicomponent de Sitter (Inflationary) Stages and the Generation of
  Perturbations,''
  JETP Lett.\  {\bf 42}, 152 (1985)
  [Pisma Zh.\ Eksp.\ Teor.\ Fiz.\  {\bf 42}, 124 (1985)].

\bibitem{Starobinsky:1982ee}
  A.~A.~Starobinsky,
  ``Dynamics Of Phase Transition In The New Inflationary Universe Scenario And
  Generation Of Perturbations,''
  Phys.\ Lett.\  B {\bf 117}, 175 (1982).


\bibitem{Linde:1993cn}
  A.~D.~Linde,
  ``Hybrid inflation,''
  Phys.\ Rev.\  D {\bf 49}, 748 (1994)
  [arXiv:astro-ph/9307002].

\bibitem{Lyth:2010ch}
  D.~H.~Lyth,
  ``Issues concerning the waterfall of hybrid inflation,''
  arXiv:1005.2461 [astro-ph.CO].

\bibitem{Lyth:2010zq}
  D.~H.~Lyth,
  ``The contribution of the hybrid inflation waterfall to the primordial
  curvature perturbation,''
  arXiv:1012.4617 [astro-ph.CO].

\bibitem{Fonseca:2010nk}
  J.~Fonseca, M.~Sasaki and D.~Wands,
  ``Large-scale Perturbations from the Waterfall Field in Hybrid Inflation,''
  JCAP {\bf 1009}, 012 (2010)
  [arXiv:1005.4053 [astro-ph.CO]].

\bibitem{Abolhasani:2010kn}
  A.~A.~Abolhasani, H.~Firouzjahi and M.~H.~Namjoo,
  ``Curvature Perturbations and non-Gaussianities from Waterfall Phase
  Transition during Inflation,''
  Class.\ Quant.\ Grav.\  {\bf 28}, 075009 (2011)
  [arXiv:1010.6292 [astro-ph.CO]].
\bibitem{Gong:2010zf}
  J.~O.~Gong and M.~Sasaki,
  ``Waterfall field in hybrid inflation and curvature perturbation,''
  JCAP {\bf 1103}, 028 (2011)
  [arXiv:1010.3405 [astro-ph.CO]].

\bibitem{Abolhasani:2012px}
  A.~A.~Abolhasani, H.~Firouzjahi, S.~Khosravi and M.~Sasaki,
  ``Local Features with Large Spiky non-Gaussianities during Inflation,''
  arXiv:1204.3722 [astro-ph.CO].



\bibitem{Chen:2009zp}
  X.~Chen and Y.~Wang,
  ``Quasi-Single Field Inflation and Non-Gaussianities,''
  JCAP {\bf 1004}, 027 (2010)
  [arXiv:0911.3380 [hep-th]].
\bibitem{Chen:2009we}
  X.~Chen and Y.~Wang,
  ``Large non-Gaussianities with Intermediate Shapes from Quasi-Single Field
  Inflation,''
  Phys.\ Rev.\  D {\bf 81}, 063511 (2010)
  [arXiv:0909.0496 [astro-ph.CO]].




\bibitem{Tolley:2009fg}
  A.~J.~Tolley and M.~Wyman,
  ``The Gelaton Scenario: Equilateral non-Gaussianity from multi-field
  dynamics,''
  Phys.\ Rev.\  D {\bf 81}, 043502 (2010)
  [arXiv:0910.1853 [hep-th]].

\bibitem{Cremonini:2010ua}
  S.~Cremonini, Z.~Lalak and K.~Turzynski,
  ``Strongly Coupled Perturbations in Two-Field Inflationary Models,''
  JCAP {\bf 1103}, 016 (2011)
  [arXiv:1010.3021 [hep-th]].

\bibitem{Achucarro:2010da}
  A.~Achucarro, J.~O.~Gong, S.~Hardeman, G.~A.~Palma and S.~P.~Patil,
  ``Features of heavy physics in the CMB power spectrum,''
  JCAP {\bf 1101}, 030 (2011)
  [arXiv:1010.3693 [hep-ph]].

\bibitem{Shiu:2011qw}
  G.~Shiu and J.~Xu,
  ``Effective Field Theory and Decoupling in Multi-field Inflation: An
  Illustrative Case Study,''
  Phys.\ Rev.\  D {\bf 84}, 103509 (2011)
  [arXiv:1108.0981 [hep-th]].

\bibitem{Cespedes:2012hu}
  S.~Cespedes, V.~Atal and G.~A.~Palma,
  ``On the importance of heavy fields during inflation,''
  arXiv:1201.4848 [hep-th].

\bibitem{Achucarro:2012sm}
  A.~Achucarro, J.~O.~Gong, S.~Hardeman, G.~A.~Palma and S.~P.~Patil,
  ``Effective theories of single field inflation when heavy fields matter,''
  arXiv:1201.6342 [hep-th].

\bibitem{Avgoustidis:2012yc}
  A.~Avgoustidis, S.~Cremonini, A.~-C.~Davis, R.~H.~Ribeiro, K.~Turzynski and S.~Watson,
  ``Decoupling Survives Inflation: A Critical Look at Effective Field Theory Violations During Inflation,''
  arXiv:1203.0016 [hep-th].

\bibitem{Assassi:2012zq}
  V.~Assassi, D.~Baumann and D.~Green,
  ``On Soft Limits of Inflationary Correlation Functions,''  arXiv:1204.4207 [hep-th].  

\bibitem{Chen:2012ge}
  X.~Chen and Y.~Wang,
  JCAP {\bf 1209}, 021 (2012)
  [arXiv:1205.0160 [hep-th]].

\bibitem{Gao:2012uq}
  X.~Gao, D.~Langlois and S.~Mizuno,
  arXiv:1205.5275 [hep-th].



\end{thebibliography}
\end{document}